# Temporal Analysis of COVID-19 Peak Outbreak

## Amit Tewari

Georgia Institute of Technology, GA, USA

Email: atewari35@gatech.edu
ORCID: https://orcid.org/0000-0002-6821-3833

*Abstract-* Intent of this research is to explore how mathematical models, specifically Susceptible-Infected-Removed (SIR) model, can be utilized to forecast peak outbreak timeline of COVID-19 epidemic amongst a population of interest starting from the date of first reported case. Till the time of this research, there was no effective and universally accepted vaccine to control transmission and spread of this infection. COVID-19 primarily spreads in population through respiratory droplets from an infected person's cough and sneeze which infects people who are in proximity. COVID-19 is spreading contagiously across the world. If health policy makers and medical experts could get early and timely insights into when peak infection rate would occur after first reported case, they could plan and optimize medical personnel, ventilators supply, and other medical resources without over-taxing the infrastructure. The predictions may also help policymakers devise strategies to control the epidemic, potentially saving many lives. Thus, it can aid in critical decision-making process by providing actionable insights into COVID-19 outbreak by leveraging available data.

## I. INTRODUCTION

Coronavirus is a large family of viruses causing illness in both animals and/or humans. Over last decade or so, several other coronaviruses are known to cause respiratory infections in humans, ranging from the common cold to more severe diseases such as Middle East Respiratory Syndrome (MERS) and Severe Acute Respiratory Syndrome (SARS). The recently discovered coronavirus causes coronavirus disease COVID-19. Today COVID-19 is causing a global pandemic affecting almost all countries [1].

Data from health organizations indicate that asymptomatic individuals can transmit the virus without themselves showing any signs of infection. Disease control organizations across the globe are investing in research on this topic and how often this happens. Recovery from novel coronavirus usually takes 14 days [2]. About 5%–15% of patients with COVID-19 infection require intensive care surveillance and ventilator support [3]. This poses a challenge for health planners and administrators as to how to optimally plan and allocate medical staff and other resources such as ventilators etc. in a large sized country such as India. According to a joint report [4] by Princeton University and The Center For Disease Dynamics, Economics & Policy (CDDEP), most of the beds and ventilators in India are concentrated in seven States only. The report also mentioned that bed capacity was saturated at hospitals and any spike in COVID-19 cases would require drastic expansion of hospital beds and ventilators. This problem represents crux of the issue that the current research is trying to address by using mathematical modeling to predict peak COVID-19 outbreak timeline in various states across India.

## II. RELATED WORK

Many previous researches and studies have attempted to employ mathematical models to provide insights into spread of influenza epidemics and pandemics [10][11][12]. Many studies have investigated historical pandemics of the 20th century [13][14][15]. Modeling techniques have also been used to understand the influence of interventions in mitigating pandemics [16].

A category of mathematical models is agent-based models (ABM) which represent a relatively recent approach to model complexities in a system composed of agents whose actions are described using simple rules. It is different from classical SIR mathematical models (which assume homogeneous population), as agent-based models try to simulate individuals with distinct characteristics and in theory can provide more realistic results. A recent study used agent-based model to evaluate COVID-19 transmission risks in facilities [18]. However, there are several difficulties associated with creating ABMs such as integration with too many features, choice of model parameters, model results being either trivial or too complex [19].

The spread of COVID-19 in India has been investigated in many researches including [20][21][22][23], but they laid little emphasis on post-model validation for peak COVID-19 timeline forecast.

With this in mind, SIR model is explored in current research to forecast peak COVID-19 outbreak over a large population in India. The SIR model was chosen because of its simplicity as well as minimalist compute and data requirements as compared to agent-based models.

## III. METHODOLOGY

### A. Model

For the purpose of this research, compartmental class of mathematical models is used in modelling COVID-19. Specifically, Kermack-McKendrick Susceptible-Infected-

Removed (SIR) model [5] is employed which distributes population in 3 compartments with labels - S, I, or R at any point of time. S is the number of susceptible individuals, I is the number of individuals infected, R is the number of individuals who have recovered and developed immunity to the infection. The number of S, I and R individuals may vary over time, but total population remains constant. Model computes the predicted number of people infected with a contagious illness in a closed population *over time*. The model assumes fixed size homogeneous population with no social or spatial structure. An individual with COVID-19 is infectious for approximately 14 days [6]. Let's assume during these 2 weeks period, they can potentially pass COVID-19 to approximately 5 people. These 2 parameters determine the model inputs viz γ, the recovery rate (= 14 days) and β, rate of infection (= 1/5 = 0.2).

Using these parameters, the time to reach peak COVID-19 outbreak starting from first reported case is predicted by solving below system of three linked nonlinear ordinary differential equations in Python [9]:

$$\frac{dS}{dt} = -\frac{\beta SI}{N}$$

$$\frac{dI}{dt} = \frac{\beta SI}{N} - \gamma I$$

$$\frac{dR}{dt} = \gamma I$$

B. Data

COVID-19 statistics data till 15-August-2020 used in this research has been sourced from NDTV [7]. Population figures for the 10 largest states in India have been taken from Statistics Times [8]. Together these 10 states constitute more than 74% of total population in India.
.

*Table 1- 10 Largest States in India by Population*

| State | Population |
|---|---|
| Uttar Pradesh | 237,882,725 |
| Bihar | 124,799,926 |
| Maharashtra | 123,144,223 |
| West Bengal | 99,609,303 |
| Madhya Pradesh | 85,358,965 |
| Rajasthan | 81,032,689 |
| Tamil Nadu | 77,841,267 |
| Karnataka | 67,562,686 |
| Gujarat | 63,872,399 |
| Andhra Pradesh | 53,903,393 |

IV. RESULTS

The SIR model predicted peak COVID-19 outbreak timelines for the states included in the research as presented in table below and corresponding figures.

*Table 2- Peak Outbreak Date Comparison (Actual [7] vs Model Predicted)*

| State | Actual Peak (Date) | Predicted Peak (Date) | Error (Days) |
|---|---|---|---|
| Uttar Pradesh | 15-Aug-2020 | 12-Aug-2020 | -3 |
| Bihar | 15-Aug-2020 | 14-Aug-2020 | -1 |
| Maharashtra | 27-Jul-2020 | 02-Aug-2020 | 6 |
| West Bengal | 10-Aug-2020 | 08-Aug-2020 | -2 |
| Madhya Pradesh | 04-Aug-2020 | 10-Aug-2020 | 6 |
| Rajasthan | 06-Aug-2020 | 22-Jul-2020 | -15 |
| Tamil Nadu | 28-Jul-2020 | 27-Jul-2020 | -1 |
| Karnataka | 03-Aug-2020 | 29-Jul-2020 | -5 |
| Gujarat | 03-Aug-2020 | 07-Aug-2020 | 4 |
| Andhra Pradesh | 01-Aug-2020 | 30-Jul-2020 | -2 |

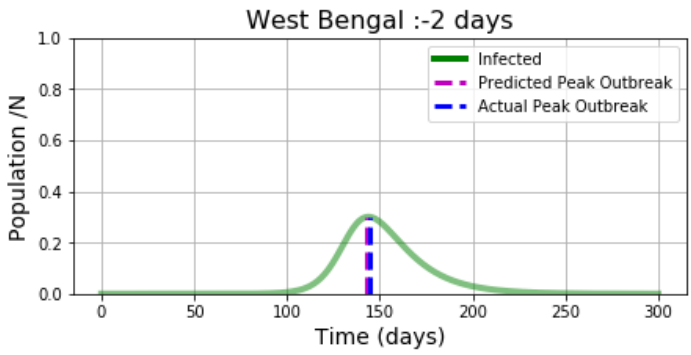
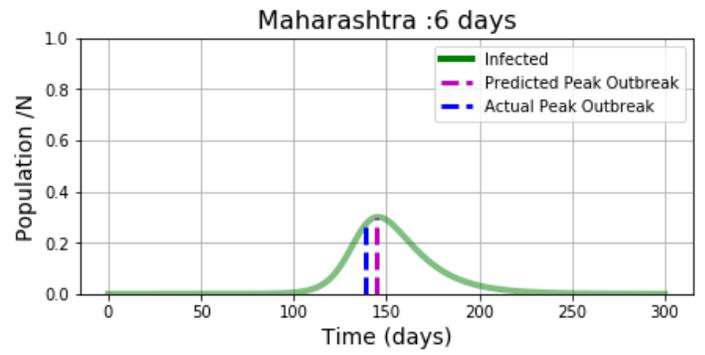
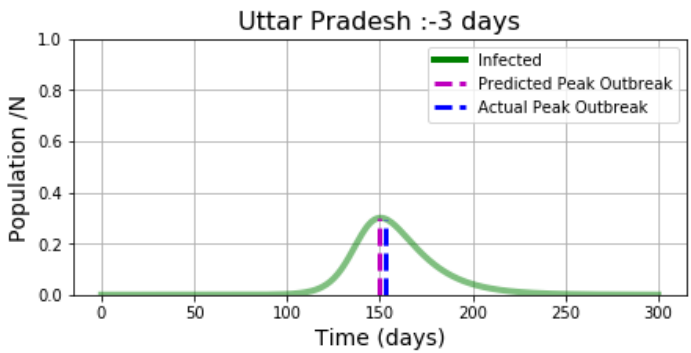
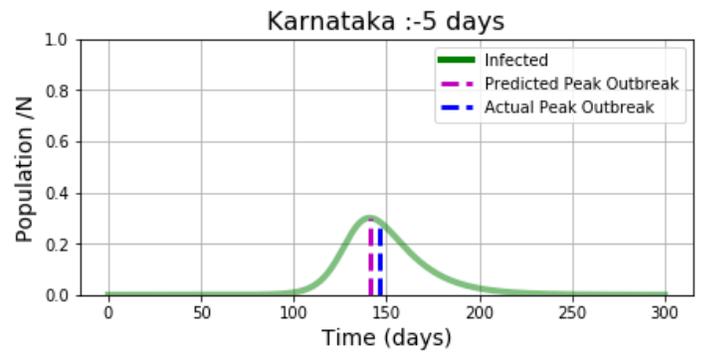
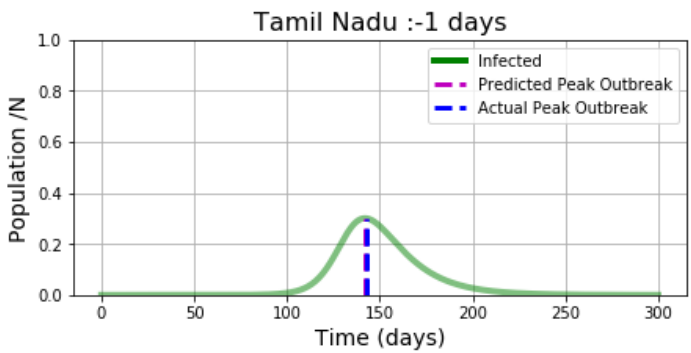
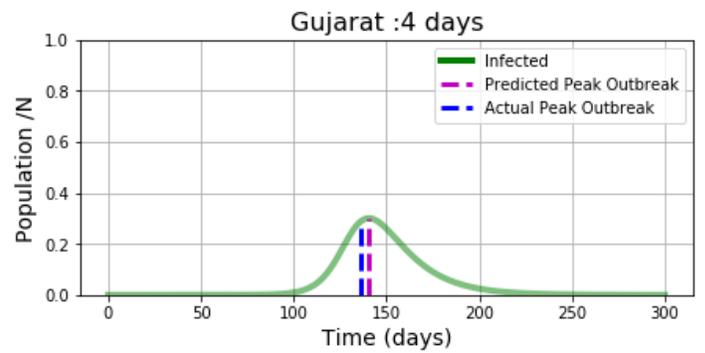
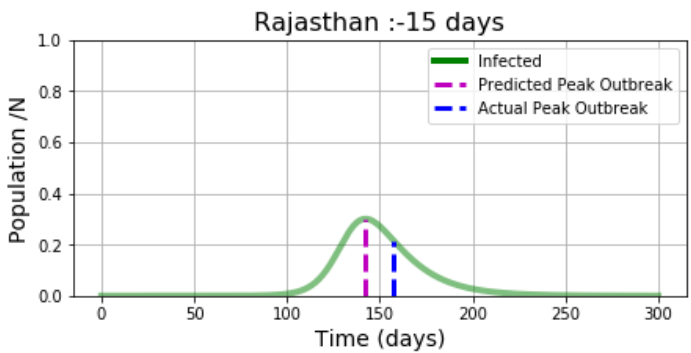
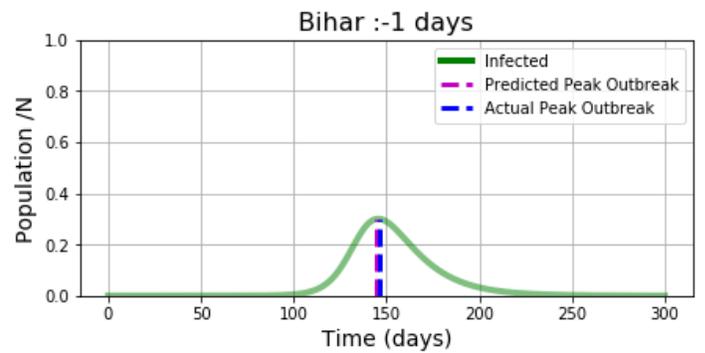
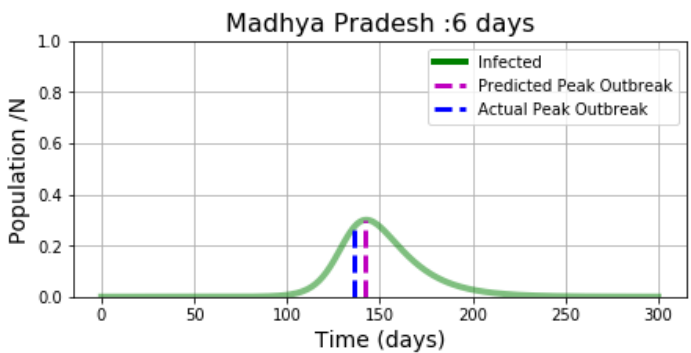
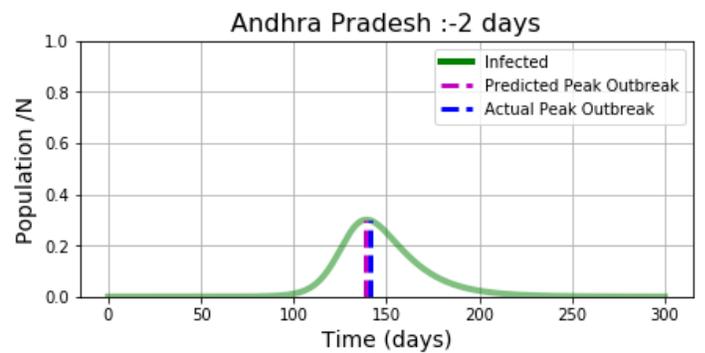

## V. DISCUSSION

This research was conducted to evaluate the feasibility of application of SIR model to predict peak COVID-19 outbreak timeline from the date of first reported case for the 10 largest states in India which together constitute more than 74% or almost 3/4th of total population in India. The broader goal is to analyze and evaluate SIR model to provide early insights to public health agencies which in turn can expedite optimum response to COVID-19 epidemic.

The research results indicate that for 9 out of these 10 states, SIR model could predict peak COVID-19 outbreak timeline from the date of first reported case with error of +/-6 days or less, Standard Deviation (SD) in error = 5.83 days and Mean Absolute Deviation (MAD) in error = 4.1 days.

## VI. CONCLUSION

Aim of this research paper was to predict COVID-19 peak timeline in various Indian states using SIR model. For 9 out of 10 largest states in India included in the research, chosen SIR model could predict peak outbreak timeline from the date of the first reported case with error of +/-6 days or less and Standard Deviation (SD) in error = 5.83 day. These 9 states constitute over 70% of total population of India.

The model results present a potential opportunity for health policy makers and medical experts to gain early and timely insights into COVID-19 peak outbreak timelines for a large proportion of population in India. They could use these insights to plan and optimize medical personnel and equipment or to devise strategies to control the epidemic, well before it hits its peak. While SIR models have been extensively used, there is little research on validating their predictions. This research provided pragmatic validation of SIR models over a large population.

## VII. LIMITATIONS

Compartmental models are in many ways favorable to other exotic models due to their simplicity and minimal computational requirements. However, SIR models assume several assumptions [17] that do not exist in real world epidemic conditions.

The SIR model assumes that there is homogeneous mixing of the infected and susceptible individuals and that the total population is constant in time. In the classic SIR model, the susceptible population decreases monotonically towards zero. However, these assumptions are not strictly valid in the case of COVID-19 outbreak, since new hot-spots spike at different times. Also, the effect of enforcing social distancing measures by respective government and health agencies has not been considered.

This research does not attempts to perform an exhaustive study because of lack of suitable data as well as uncertainty in different factors, namely, the degree of home isolation, restrictions in social contact, the initial number of infected and exposed individuals, variations in incubation and infectious periods, and the fatality rate.

## VIII. FUTURE SCOPE

Population density, geographic area, demographics such as age and effect of social isolation etc. are possibly some parameters to consider and include in building more advanced epidemiology models for predicting peak epidemic outbreak timeline. Disease spread models can also be used to predict number of infected individuals to better manage epidemics.

ACKNOWLEDGMENT

The author would like to thank the editor and the reviewers for their helpful comments and review that contributed to improve the manuscript.